\documentclass[longauth]{aa} 
\usepackage{graphicx,natbib}
\usepackage{txfonts}
\begin{document}
   \title{The correlated optical and radio variability of \object{BL Lacertae}}

   \subtitle{WEBT data analysis 1994--2005\thanks{The radio-to-optical 
   data presented in this paper are stored in the WEBT archive; for questions regarding their availability, please contact the WEBT President Massimo Villata ({\tt villata@oato.inaf.it}).} }

   \author{M.~Villata                 \inst{ 1}
   \and   C.~M.~Raiteri               \inst{ 1}
   \and   V.~M.~Larionov              \inst{ 2,3,4}
   \and   M.~G.~Nikolashvili          \inst{ 5}
   \and   M.~F.~Aller                 \inst{ 6}
   \and   U.~Bach                     \inst{ 7}
   \and   D.~Carosati                 \inst{ 8}
   \and   F.~Hroch                    \inst{ 9}
   \and   M.~A.~Ibrahimov             \inst{10}
   \and   S.~G.~Jorstad               \inst{11,2}
   \and   Y.~Y.~Kovalev               \inst{ 7,12}
   \and   A.~L\"ahteenm\"aki          \inst{13}
   \and   K.~Nilsson                  \inst{14}
   \and   H.~Ter\"asranta             \inst{13}
   \and   G.~Tosti                    \inst{15}
   \and   H.~D.~Aller                 \inst{ 6}
   \and   A.~A.~Arkharov              \inst{ 3}
   \and   A.~Berdyugin                \inst{14}
   \and   P.~Boltwood                 \inst{16}
   \and   C.~S.~Buemi                 \inst{17}
   \and   R.~Casas                    \inst{18}
   \and   P.~Charlot                  \inst{19,20}
   \and   J.~M.~Coloma                \inst{21}
   \and   A.~Di~Paola                 \inst{22}
   \and   G.~Di~Rico                  \inst{23}
   \and   G.~N.~Kimeridze             \inst{ 5}
   \and   T.~S.~Konstantinova         \inst{ 2}
   \and   E.~N.~Kopatskaya            \inst{ 2}
   \and   Yu.~A.~Kovalev              \inst{12}
   \and   O.~M.~Kurtanidze            \inst{ 5}
   \and   L.~Lanteri                  \inst{ 1}
   \and   E.~G.~Larionova             \inst{ 2}
   \and   L.~V.~Larionova             \inst{ 2}
   \and   J.-F.~Le~Campion            \inst{19,20}
   \and   P.~Leto                     \inst{24}
   \and   E.~Lindfors                 \inst{14}
   \and   A.~P.~Marscher              \inst{11}
   \and   K.~Marshall                 \inst{25}
   \and   J.~P.~McFarland             \inst{25}
   \and   I.~M.~McHardy               \inst{26}
   \and   H.~R.~Miller                \inst{25}
   \and   G.~Nucciarelli              \inst{15}
   \and   M.~P.~Osterman              \inst{25}
   \and   M.~Pasanen                  \inst{14}
   \and   T.~Pursimo                  \inst{27}
   \and   J.~A.~Ros                   \inst{21}
   \and   A.~C.~Sadun                 \inst{28}
   \and   L.~A.~Sigua                 \inst{ 5}
   \and   L.~Sixtova                  \inst{ 9}
   \and   L.~O.~Takalo                \inst{14}
   \and   M.~Tornikoski               \inst{13}
   \and   C.~Trigilio                 \inst{17}
   \and   G.~Umana                    \inst{17}
   \and   G.~Z.~Xie                   \inst{29,30}
   \and   X.~Zhang                    \inst{30}
   \and   S.~B.~Zhou                  \inst{30}
 }

   \offprints{M.\ Villata}

   \institute{
          INAF, Osservatorio Astronomico di Torino, Italy                                                     
   \and   Astronomical Institute, St.-Petersburg State University, Russia                                     
   \and   Pulkovo Observatory, Russia                                                                         
   \and   Isaac Newton Institute of Chile, St.-Petersburg Branch, Russia                                      
   \and   Abastumani Astrophysical Observatory, Georgia                                                       
   \and   Department of Astronomy, University of Michigan, MI, USA                                            
   \and   Max-Planck-Institut f\"ur Radioastronomie, Germany                                                  
   \and   Armenzano Astronomical Observatory, Italy                                                           
   \and   Institute of Theoretical Physics and Astrophysics, Masaryk University, Czech Republic               
   \and   Ulugh Beg Astronomical Institute, Academy of Sciences of Uzbekistan, Uzbekistan                     
   \and   Institute for Astrophysical Research, Boston University, MA, USA                                    
   \and   Astro Space Center of Lebedev Physical Institute, Russia                                            
   \and   Mets\"ahovi Radio Observatory, Helsinki University of Technology TKK, Finland                       
   \and   Tuorla Observatory, Department of Physics and Astronomy, University of Turku, Finland               
   \and   Dipartimento di Fisica, Universit\`a di Perugia, Italy                                              
   \and   Boltwood Observatory, ON, Canada                                                                    
   \and   INAF, Osservatorio Astrofisico di Catania, Italy                                                    
   \and   Institut de Ci\`encies de l'Espai (IEEC-CSIC), Barcelona, Spain                                     
   \and   Universit\'e de Bordeaux, Observatoire Aquitain des Sciences de l'Univers, France                   
   \and   CNRS, Laboratoire d'Astrophysique de Bordeaux -- UMR5804, France                                    
   \and   Agrupaci\'o Astron\`omica de Sabadell, Spain                                                        
   \and   INAF, Osservatorio Astronomico di Roma, Italy                                                       
   \and   INAF, Osservatorio Astronomico di Collurania Teramo, Italy                                          
   \and   INAF, Istituto di Radioastronomia, Sezione di Noto, Italy                                           
   \and   Department of Physics and Astronomy, Georgia State University, GA, USA                              
   \and   Department of Physics and Astronomy, University of Southampton, UK                                  
   \and   Nordic Optical Telescope, Roque de los Muchachos Astronomical Observatory, TF, Spain                
   \and   Department of Physics, University of Colorado Denver, CO, USA                                       
   \and   National Astronomical Observatories/Yunnan Observatory, Chinese Academy of Sciences, China          
   \and   Physics Department, Yunnan University, China                                                        
 }

   \date{}

 
  \abstract
   {Since 1997, BL Lacertae has undergone a phase of high optical activity,
with the occurrence of several prominent outbursts. 
Starting from 1999, the Whole Earth Blazar Telescope (WEBT) consortium has organized
various multifrequency campaigns on this blazar, collecting tens of thousands of data points.
One of the main issues in the study of this huge dataset has been the search for correlations
between the optical and radio flux variations, and for possible periodicities in the light curves.
The analysis of the data assembled during the first four campaigns (comprising also archival data to cover
the period 1968--2003) revealed a fair optical-radio correlation in 1994--2003, with a delay of the hard radio events of $\sim 100$ days. Moreover, various statistical methods suggested the existence of a radio periodicity of $\sim 8$ years.}
   {In 2004 the WEBT started a new campaign to extend the dataset to the most recent observing seasons, 
in order to possibly confirm and better understand the previous results.}
   {In this campaign we have collected and assembled about 11000 new optical observations from twenty
telescopes, plus near-IR and radio data at various frequencies. 
Here, we perform a correlation analysis on the long-term $R$-band and radio light curves.}
   {In general, we confirm the $\sim 100$-day delay of the hard radio events with respect to the 
optical ones, even if longer ($\sim 200$--300 days) time lags are also found in particular periods.
The radio quasi-periodicity is confirmed too, but the ``period" seems to progressively lengthen from 7.4 to 9.3 years in the last three cycles. The optical and radio behaviour in the last forty years suggests
a scenario where geometric effects play a major role. In particular, the alternation of enhanced and suppressed optical activity (accompanied by hard and soft radio events, respectively) can be explained in terms of an emitting plasma flowing along a rotating helical path in a curved jet.}
   {}


   \maketitle

\section{Introduction}

BL Lacertae ($z=0.0688 \pm 0.0002$; \citealt{mil77}) is the prototype of a class of active galactic nuclei (AGNs), the BL Lac objects, which are well known for their pronounced variability at all wavelengths, from the radio to the $\gamma$-ray band (see e.g.\ \citealt{vil02} for an extended description of this object). It has been one of the favourite targets of the Whole Earth Blazar Telescope 
(WEBT)\footnote{{\tt http://www.oato.inaf.it/blazars/webt/}}, which has organized several multifrequency campaigns on this source. 
Two short campaigns took place in 1999, in conjunction with observations of the X-ray satellites ASCA and BeppoSAX \citep{rav02}. 
The campaign carried out in May 2000 -- January 2001 was particularly successful, with periods of exceptionally dense sampling that allowed \citet{vil02} to follow the intranight flux variations in detail. The optical spectrum was found to be weakly sensitive to the long-term brightness trend, but to strictly follow the short-term flux variations, becoming bluer when brighter. The authors suggested that the
essentially achromatic modulation of the flux base level on long time scales is due to a variation of the relativistic Doppler beaming factor, and that this variation is likely due to a change of the viewing angle. In contrast, the strongly chromatic fast variability may be the result of intrinsic energetic processes. 
The data of this WEBT campaign on BL Lac were compared to the X-ray data taken in the same period by BeppoSAX and RXTE by \citet{boe03}, who found some hint of correlation.

A subsequent campaign was carried out in 2001--2003. On that occasion, historical optical and radio light curves were reconstructed back to 1968, and both colour and time series analyses were performed. 
Colour analysis on a longer time period confirmed the conclusions by \citet{vil02}, and allowed \citet{vil04a} to quantify the degree of chromatism of both the short-term and long-term variability components.
The optical spectral behaviour was further analysed by \citet{pap07}, who interpreted 
the bluer-when-brighter mild chromatism of the long-term
variations in terms of Doppler factor variations due to changes in the
viewing angle of a curved and inhomogeneous jet.

The time series analysis performed by \citet{vil04b} showed that the main radio outbursts repeat every $\sim 8$ years \citep[see also][]{cia04}, with a possible progressive stretching of the period.
Moreover, when considering the best-sampled time interval 1994--2003, the optical light curve was found to correlate with the radio hardness ratios, with a radio time delay of about 100 days.
Thus, \citet{vil04b} proposed a scenario where the variability mechanism 
propagates downstream in the jet, crossing less and less opaque regions.
In the inner regions it produces connected optical and hard radio events, while in outer zones it gives rise to softer events that are apparently uncorrelated with the former ones.
The WEBT light curves were then used by \citet{bac06} for a comparison with the time evolution of the VLBI core and jet structure on parsec and subparsec scales. 
The authors showed that sometimes prominent jet features can become as bright as, or even brighter than, the core, which complicates the cross-correlation analysis between the radio light curves and those at other frequencies. However, they found that the radio hardness ratios derived from the WEBT light curves can trace the variability of the VLBI core. Their analysis then confirmed the optical-radio time lags of \citet{vil04b}. Jet bending was invoked to explain both the modest optical variability in the period 1981--1996 with suppressed radio-optical correlation, and the low-emission region in the jet around 1 mas from the core.

In this paper we present optical ($R$ band) and radio data taken during the WEBT campaign covering the 2002--2005 period\footnote{The radio-to-optical data presented in this paper are stored in the WEBT archive; for questions regarding their availability, please contact the WEBT President Massimo Villata ({\tt villata@oato.inaf.it}). More information at {\tt http://www.oato.inaf.it/blazars/webt/}.}. Their addition to the historical light curves allows us to further investigate both the radio quasi-periodicity and the correlation between optical flux and radio hardness ratio found by \citet{vil04b}.

A new campaign on BL Lacertae was organized by the WEBT in the 2007--2008 observing season, including three pointings by the XMM-Newton satellite as well as spectroscopic observations with the Telescopio Nazionale Galileo (TNG). The results of this further WEBT effort on BL Lac will be presented elsewhere (Raiteri et al., in preparation).

\section{Observations and results}

   \begin{figure*}
   \centering
   \includegraphics{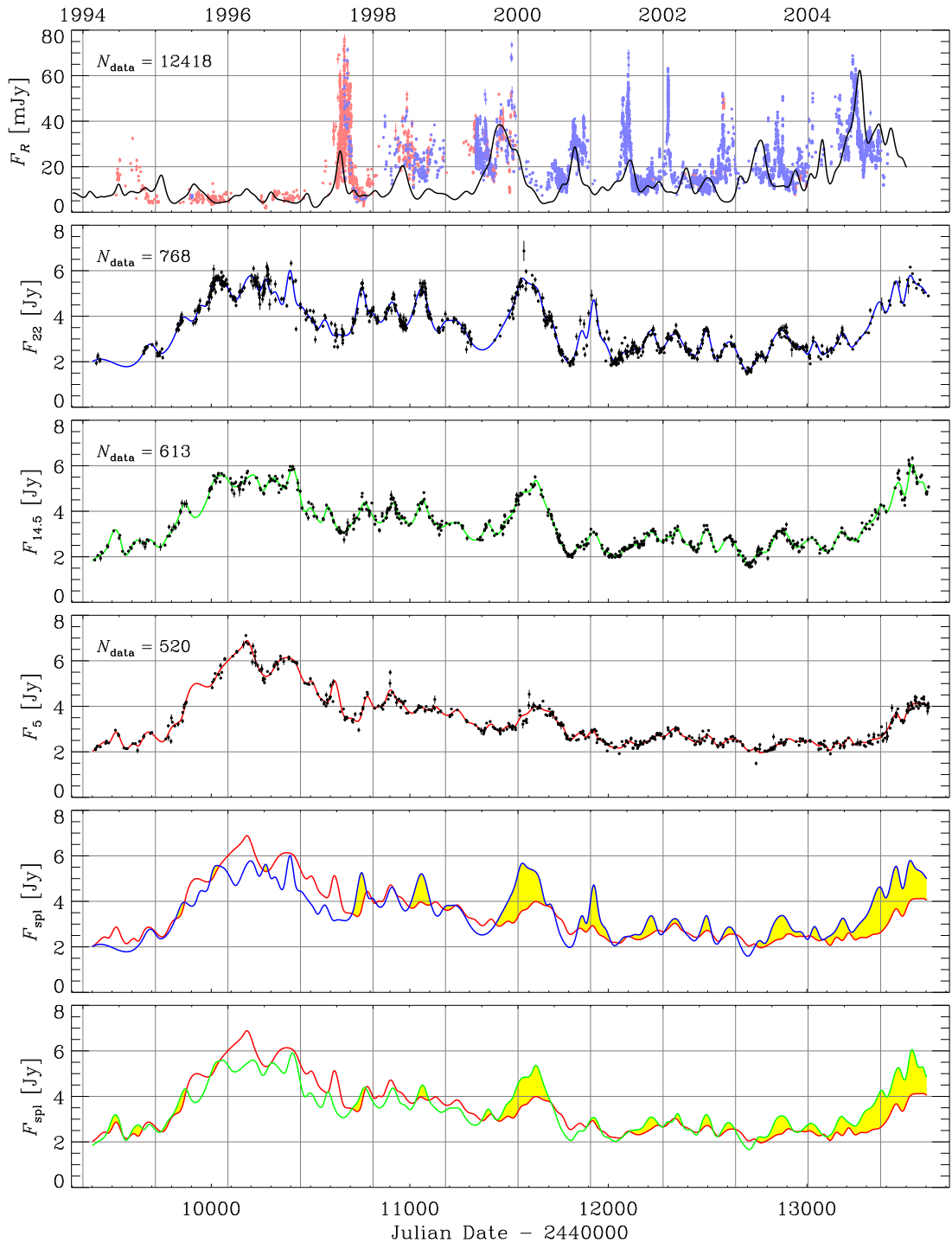}
   \caption{$R$-band flux-density light curve (top, mJy) and radio light curves (Jy)
   at 22, 14.5, and 5 GHz with their cubic spline interpolations underplotted;
   grey horizontal and vertical lines have been drawn to guide the eye 
   through the flux variations;
   the fifth and sixth panels show the 22 GHz (blue line) and the 14.5 GHz (green line)
   splines compared to the 5 GHz (red) one; the regions where the higher-frequency 
   spline exceeds the lower-frequency one have been highlighted in yellow.
   In the top panel the solid line represents the average between $H_{22}$ and $H_{14}$,
   $H_{\rm mean}$, raised to 3.5, multiplied by 10, and shifted in time by $-100$ days,
   for an easier comparison with the optical light curve.}
    \label{radop}
    \end{figure*}

During the 2002--2005 WEBT campaign, we assembled about 11000 observations in the optical $UBVRI$ bands from 16 observatories (20 telescopes); in order of longitude they are:
Yunnan, 
Mt. Maidanak, 
Abastumani, 
Crimean,
Tuorla, 
MonteBoo, 
Armenzano, 
Perugia, 
Torino, 
Sabadell, 
Bordeaux, 
Roque de los Muchachos (KVA, Liverpool, NOT, and TNG),
Boltwood, 
Capilla Peak, 
Lowell (Hall and Perkins), 
and Kitt Peak (SARA). 
$JHK$ data were collected from 3 observatories: Campo Imperatore, TIRGO, and Roque de los Muchachos (NOT). Radio observations were carried out between 1 and 43 GHz at: SAO RAS (RATAN-600)\footnote{See e.g.\ \citet{kor79}; \citet{kov99}.}, Mets\"ahovi\footnote{See e.g.\ \citet{ter05} and references therein.}, Noto,  Medicina\footnote{See e.g.\ \citet{bac07}.}, MDSCC (PARTNeR), and UMRAO.
Optical and near-IR data were processed as explained in \citet{vil02}. 
Radio data were collected as already calibrated flux densities and complemented by data from the VLA/VLBA Polarization Calibration Database\footnote{\tt http://www.vla.nrao.edu/astro/calib/polar/}.

The top panel of Fig.\ \ref{radop} shows the optical $R$-band flux-density light curve starting from 1994 (12418 data points). It was built with data collected by the WEBT in five campaigns (blue dots), including the 2002--2005 one, as well as with literature data\footnote{Besides the papers already cited in \citet{vil04a}, literature data were taken from \citet{hag04}, \citet{zha04}, and \citet{gu06}.} (red dots). The flux densities have been corrected for the Galactic extinction and the host galaxy contribution has been subtracted according to \citet[][see also \citealt{vil04a}]{vil02}.
Radio light curves at 22, 14.5, and 5 GHz are shown in the following three panels, together with cubic spline interpolations through the 30-day binned data.
These splines are reported in the last two panels, 
to compare the 22 and 14.5 GHz variations to the 5 GHz ones.
Following \citet{vil04b}, we define the hardness ratio $H_{22}$ as the ratio between the 22 and 5 GHz flux densities. Analogously, $H_{14}$ is the ratio between the 14.5 and 5 GHz flux densities. The events where these hardness ratios are greater than 1 are highlighted in yellow and called ``hard" events. ``Soft" events are characterized by higher flux densities at lower frequencies.

After a period of moderate activity \citep[1981--1996; see Fig.\ 1 in][]{vil04b}, from 1997 the source has shown several prominent outbursts in the optical band.
In contrast, the radio activity in the last 25 years does not present a corresponding change of behaviour.
In particular, the major radio outburst of 1995--1996 occurred during a ``quiescent" optical phase, and it was stronger at longer wavelengths. In the following years the reverse was observed, with outbursts appearing enhanced at the higher radio frequencies. 
This behaviour led \citet{vil04b} to distinguish the soft radio events, with no optical counterpart, from the hard events, which are correlated with the optical outbursts.
Indeed, a cross-correlation analysis with the discrete correlation function \citep[DCF; see][]{ede88,huf92,pet98} method yielded a fair correlation ($\rm DCF \sim 0.5$) between the optical fluxes and $H_{22}$, with a time delay of the latter of about 100 days.

\begin{figure}
   \resizebox{\hsize}{!}{\includegraphics{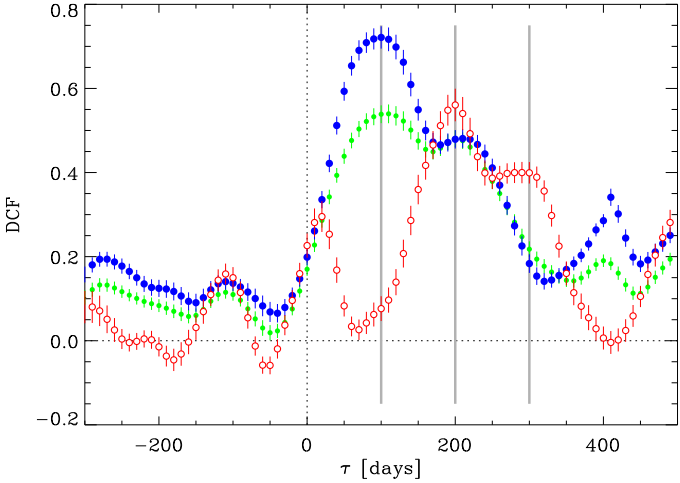}}
   \caption{Discrete correlation functions between the optical flux densities and the radio hardness ratio
$H_{\rm mean}$ (see text) for different optical time intervals: total 1994--2005 light curve (green dots), total period excluding the interval $\rm JD=2452150$--2453200 (blue filled circles), and this
interval alone (red empty circles). The grey vertical bars indicate the typical time delays of 100, 200, and 300 days.}
    \label{dcf}
    \end{figure}

One of the main purposes of the new WEBT campaign in 2002--2005 was to check if this 
result would have been confirmed by the data of the new period.
In the top panel of Fig.\ \ref{radop} the solid line represents the average between $H_{22}$ and $H_{14}$, $H_{\rm mean}$, shifted in time by $-100$ days and properly scaled.
It allows us to recognize that most optical outbursts have their counterpart in the $H_{\rm mean}$ curve, but with a less clear correspondence in the period $\sim 2002$--2004.5.
When cross-correlating the optical light curve with $H_{\rm mean}$, we obtain a result (green dots in Fig.\ \ref{dcf}) very similar to that of \citet{vil04b}, with the DCF peak at $\sim 100$ days slightly enhanced ($\sim 0.55$), and a secondary peak at about 200 days now more pronounced.
Hence, it seems that the hard radio events can lag behind the optical ones with two different time delays.
To see whether a change of behaviour occurred at some stage, we divided the optical light curve in two periods. The first period runs from 1994 to 2005, with the exclusion of the time interval where we noticed a lack of correspondence between optical and radio events, i.e.\ $\rm JD=2452150$--2453200 ($\sim 2001.7$--2004.5), which represents the second period.
The blue filled circles in Fig.\ \ref{dcf} show the DCF between the optical data in the first period and $H_{\rm mean}$. The peak at $\sim 100$ days appears strongly enhanced ($\ga 0.7$), while the second one is essentially unchanged. 
On the other hand, the red empty circles show the radio delays in the second period: the main peaks are now at about 200 and 300 days.
Thus, the hard radio outbursts seem to lag behind the optical ones by about 100 days in 1997--2001 with a possible lengthening to $\sim 200$ days, while later the delay appears to increase in the range 200--300 days, until it comes back to the original value in 2004--2005.

Another issue to be investigated is the quasi-periodicity ($\sim 8$ years) of the main radio outbursts, particularly evident in the longer-wavelength light curves \citep[see][]{vil04b}.
This result is based on the presence of three events peaking around 1980, 1988, and 1996.
Moreover, \citet{vil04b} also noticed a possible stretching of the period: the separation between the first two outbursts is noticeably less than 8 years, and the following time interval is slightly longer than 8 years. The occurrence of the radio outburst observed in 2005 seems to confirm an increasing time separation between the events, the last time interval being of at least 9 years (see Fig.\ \ref{radop}).

\begin{figure}
   \resizebox{\hsize}{!}{\includegraphics{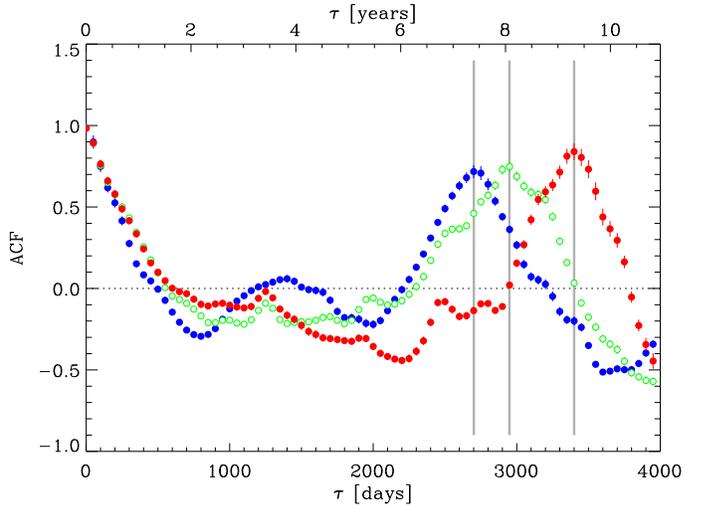}}
   \caption{Auto-correlation functions on subsequent pairs of radio outbursts at 8 GHz.
Blue filled circles refer to the period $\rm JD = 2443000$--2449000, green empty circles to $\rm JD = 2446000$--2452000, and red filled circles to $\rm JD = 2449000$--2454000.}
    \label{acf}
    \end{figure}

To quantify the effect, we auto-correlated the 8 GHz light curve (the best sampled one) by means of the DCF, taking the main radio outbursts two by two. In Fig.\ \ref{acf} we show the results for the time intervals $\rm JD = 2443000$--2449000 ($\sim 1976.6$--1993.0; blue filled circles), $\rm JD = 2446000$--2452000 ($\sim 1984.8$--2001.2; green empty circles), and $\rm JD = 2449000$--2453700 ($\sim 1993.0$--2005.9; red filled circles).
By considering the centroid of the peak distribution \citep[see e.g.][]{rai03}, the first two outbursts are found to be separated by 2701 days ($\sim 7.4$ years), the second pair by 2949 days ($\sim 8.1$ years), and the last events present a time lag of 3400 days ($\sim 9.3$ years).

\section{Discussion}

In this paper we have analysed new (unpublished) as well as old multifrequency data collected by the WEBT consortium during various campaigns on BL Lacertae to address two main issues: i) the correlation and time delay between the optical and radio flux variations and ii) the possible quasi-periodicity of the main radio outbursts.

We have found that the hard radio events lag behind the optical outbursts with time scales varying from about 100 to 300 days. In particular, the longest delays occurred in the period 2002--2004.5.
A possible explanation for this variable radio delay can be found in an inhomogeneous jet scenario, where the mechanism producing the flux enhancement propagates downstream, triggering optical and then radio outbursts as it crosses the corresponding emitting regions. If the jet changes its orientation with respect to the line of sight, relativistic effects produce a variation in the observed time scales. In particular, if the jet portion emitting the optical-to-radio radiation has a smaller (larger) viewing angle, this yields a shorter (longer) delay of the radio events with respect to the optical ones. 
However, the optical light curve in the period 2002--2004.5 does not show any peculiar behaviour that may suggest a significant change of the bulk orientation of the optical emitting region.
Hence, the change in orientation must have occurred outward in the jet, toward the radio region, i.e.\ the jet changed its curvature. Indeed, by looking at the radio light curves in Fig.\ \ref{radop}, we can see that the minimum radio level is achieved just around 2003, and the following hard radio outburst of the same year is the most delayed one with respect to its optical counterpart of late 2002.

Auto-correlation analysis performed on the historical 8 GHz light curve revealed a progressive lengthening of the time lag between subsequent major radio outbursts: 7.4, 8.1, and 9.3 years. If we assume that the intrinsic period is constant, the observed stretching might be due to increased light travel time. Even considering that the distance between the emitting region and the AGN core does not vary, a change in the orientation of the former departing from the AGN line of sight would augment the distance to us and hence the light travel time.
High-resolution VLBA maps of BL Lacertae at 43 GHz presented by \citet{mar08} show that the resolved jet structure may have typical length scales of the order of a light year. These distances may imply much longer de-projected lengths, so that our hypothesis of a $\sim 1$-year increase of the light travel time appears to be reasonable.

   \begin{figure*}
   \sidecaption
   \includegraphics[angle=-90,width=13cm,clip]{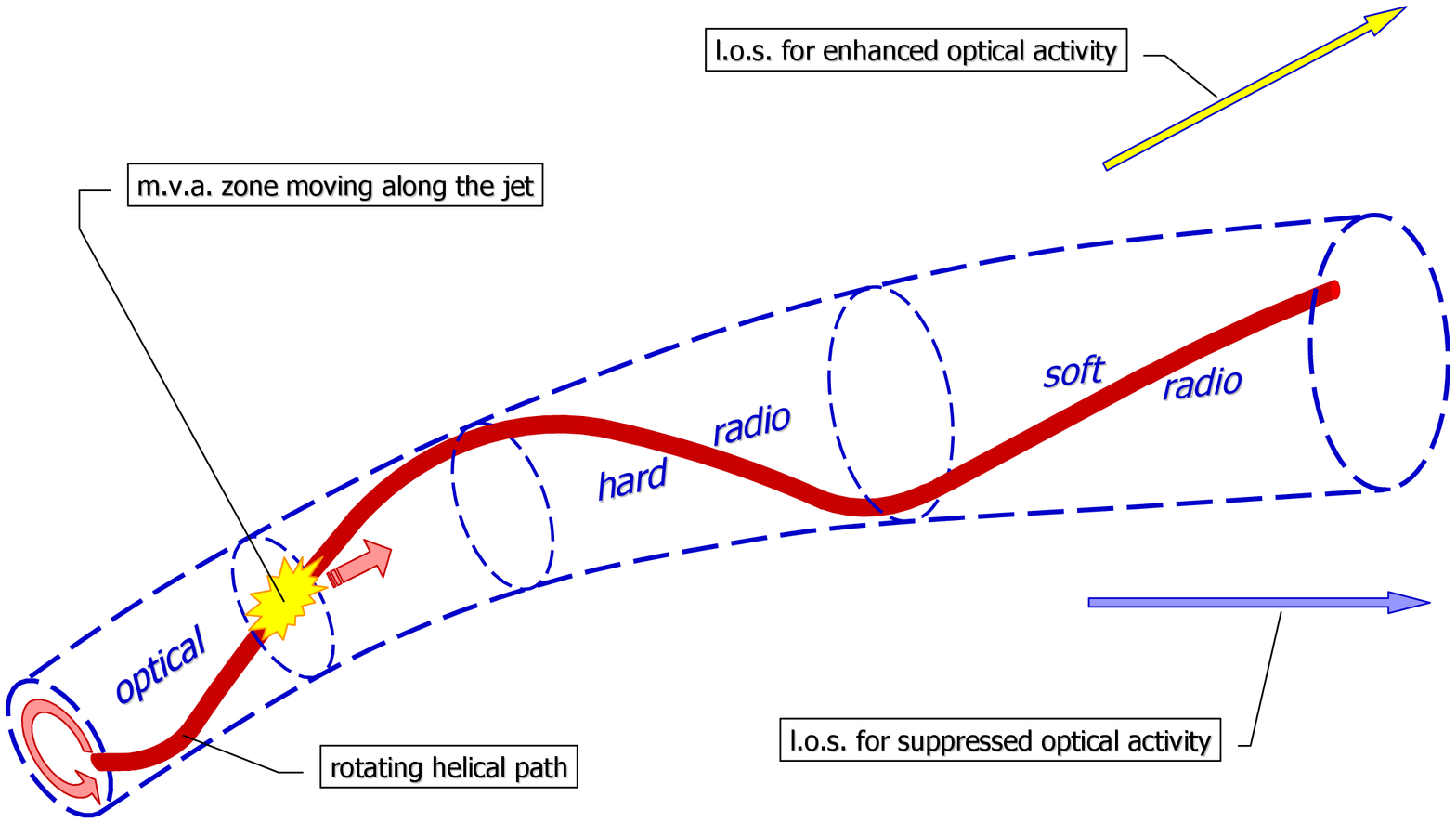}
   \caption{Sketch of a curved jet model with a rotating helical path.}
    \label{jet}
    \end{figure*}

As a general view of the BL Lac multiwavelength emission behaviour in the last about forty years, we can distinguish three different phases. In 1968--1980 the source appears to be active in both the optical and radio bands, and this phase culminates and ends with the exceptional 1980--1981 radio outburst, which presented a particularly hard spectrum in its first major peak \citep[see Fig.\ 1 in][]{vil04b}. In the second phase, 1981--1996, the reverse is observed: low optical levels and variability accompanied by a reduced and softer radio activity, which however culminates in the large, soft radio outburst of 1995--1997. The third phase seems more similar to the first one, with strongly enhanced optical activity and hard radio events. This behaviour may be explained by an orientation change of the emitting jet, where the optical and hard-radio emitting regions are better aligned with the line of sight during the first and third phases, the soft-radio part of the jet having a larger viewing angle. The opposite for the second phase. 

In addition to this scenario, we also observe the radio ``periodicity", where the major periodic events maintain the hard or soft nature of the phase they belong to. In other words, whatever the periodicity mechanism is, the 1980--1981 outburst occurred when the radio region was very well aligned, especially in the harder part. During the second, 1987--1988 outburst, the radio (especially hard) region was probably more poorly aligned, while the very soft 1995--1997 event occurred when the lower-frequency emitting regions had a smaller viewing angle. The last, hard radio outburst is located well inside the hard last phase. The radio periodicity we observe is much better seen at longer radio wavelengths, which are less affected by the ``noise" of the optical and hard-radio activity, so that the relative amplitude of the four major outbursts is probably essentially due to the corresponding emitting region orientation. 

Inside this long-term behaviour, probably of geometric origin, one can wonder what the nature of the single events is, i.e.\ whether the optical and correlated radio events shown in Fig.\ \ref{radop} are due to intrinsic or geometric variations. \citet{vil04a} argued that the poor chromatism of the optical outbursts compared to the strong chromatism of the fast flares suggests a geometric origin of the former. 
In this view, the mechanism producing the flux enhancement that travels downstream would be the Doppler boosting arising from a rotating helical path \citep[see e.g.][]{vil99,ost04}. The portion of the helical path having the smallest viewing angle successively crosses and lights up the different emitting regions. Alternatively, the existence of physical disturbances going through the jet may also explain the observed correlated events, but in this case the above-mentioned different chromatism does not have a simple explanation.

Figure \ref{jet} helps clarify the ``geometrical" model, sketching the part of the jet emitting optical-to-radio frequencies. The emitting plasma flows along a rotating helical path. 
The yellow spot marks the portion of the helical path having the minimum viewing angle when the line of sight is represented by the yellow arrow. As the helix rotates, this zone shifts along the jet, thus originating optical and then radio outbursts when crossing the corresponding emitting regions. The orientation of the curved jet with respect to the yellow line of sight yields a Doppler enhancement of the optical activity. In contrast, when the jet moves and the line of sight is represented by the blue arrow, the alignment of the optical emitting region changes and consequently the optical activity is suppressed. At the same time the soft-radio emitting region now has a smaller viewing angle and hence the soft-radio activity increases.
This bulk jet motion would thus explain the alternate epochs of optical (and hard-radio) and soft-radio activity noticed in the historical light curves.

\begin{acknowledgements}
The St.\ Petersburg team acknowledges support from Russian Foundation for Basic Researches, grant 09-02-00092.
AZT-24 observations at Campo Imperatore are made within an agreement between Pulkovo, Rome and Teramo observatories.
This research has made use of data from the University of Michigan Radio Astronomy Observatory,
which is supported by the National Science Foundation and by funds from the University of Michigan.
This work is partly based on observations with the Medicina and Noto radio telescopes operated by INAF -- Istituto di Radioastronomia.
The Liverpool Telescope is operated on the island of La Palma by Liverpool John Moores University in the Spanish Observatorio del Roque de los Muchachos of the Instituto de Astrof\'{\i}sica de Canarias with financial support from the UK Science and Technology Facilities Council.
Y.\ Y.\ Kovalev is a Research Fellow of the Alexander von Humboldt 
Foundation. \mbox{RATAN--600} observations are partly supported by the 
Russian Foundation for Basic Research (projects 01-02-16812, 
05-02-17377, 08-02-00545).
The Mets\"ahovi team acknowledges the support from the Academy of Finland.
This work is partly based on observations made with the Nordic Optical Telescope, operated
on the island of La Palma jointly by Denmark, Finland, Iceland, Norway, and Sweden, in the Spanish Observatorio del Roque de los Muchachos of the Instituto de Astrof\'{\i}sica de Canarias. 
This work is partly based on observations made with the Italian Telescopio Nazionale Galileo (TNG) operated on the island of La Palma by the Fundaci\'on Galileo Galilei of the INAF (Istituto Nazionale di Astrofisica) at the Spanish Observatorio del Roque de los Muchachos of the Instituto de Astrof\'{\i}sica de Canarias.
\end{acknowledgements}


\begin{thebibliography}{22}
\expandafter\ifx\csname natexlab\endcsname\relax\def\natexlab#1{#1}\fi

\bibitem[{{Bach} {et~al.}(2007){Bach}, {Raiteri}, {Villata}, {Fuhrmann},
  {Buemi}, {Larionov}, {Letog}, {Arkharov}, {Coloma}, {di Paola}, {Dolci},
  {Efimova}, {Forn{\'e}}, {Ibrahimov}, {Hagen-Thorn}, {Konstantinova},
  {Kopatskaya}, {Lanteri}, {Kurtanidze}, {Maccaferri}, {Nikolashvili},
  {Orlati}, {Ros}, {Tosti}, {Trigilio}, \& {Umana}}]{bac07}
{Bach}, U., {Raiteri}, C.~M., {Villata}, M., {et~al.} 2007, \aap, 464, 175

\bibitem[{{Bach} {et~al.}(2006){Bach}, {Villata}, {Raiteri}, {Agudo}, {Aller},
  {Aller}, {Denn}, {G{\'o}mez}, {Jorstad}, {Marscher}, {Mutel}, \&
  {Ter{\"a}sranta}}]{bac06}
{Bach}, U., {Villata}, M., {Raiteri}, C.~M., {et~al.} 2006, \aap, 456, 105

\bibitem[{{B{\"o}ttcher} {et~al.}(2003){B{\"o}ttcher}, {Marscher}, {Ravasio},
  {Villata}, {Raiteri}, {Aller}, {Aller}, {Ter{\"a}sranta}, {Mang},
  {Tagliaferri}, {Aharonian}, {Krawczynski}, {Kurtanidze}, {Nikolashvili},
  {Ibrahimov}, {Papadakis}, {Tsinganos}, {Sadakane}, {Okada}, {Takalo},
  {Sillanp{\"a}{\"a}}, {Tosti}, {Ciprini}, {Frasca}, {Marilli}, {Robb},
  {Noble}, {Jorstad}, {Hagen-Thorn}, {Larionov}, {Nesci}, {Maesano},
  {Schwartz}, {Basler}, {Gorham}, {Iwamatsu}, {Kato}, {Pullen},
  {Ben{\'{\i}}tez}, {de Diego}, {Moilanen}, {Oksanen}, {Rodriguez}, {Sadun},
  {Kelly}, {Carini}, {Miller}, {Catalano}, {Dultzin-Hacyan}, {Fan},
  {Ghisellini}, {Ishioka}, {Karttunen}, {Kein{\"a}nen}, {Kudryavtseva},
  {Lainela}, {Lanteri}, {Larionova}, {Matsumoto}, {Mattox}, {McHardy},
  {Montagni}, {Nucciarelli}, {Ostorero}, {Papamastorakis}, {Pasanen},
  {Sobrito}, \& {Uemura}}]{boe03}
{B{\"o}ttcher}, M., {Marscher}, A.~P., {Ravasio}, M., {et~al.} 2003, \apj, 596,
  847

\bibitem[{{Ciaramella} {et~al.}(2004){Ciaramella}, {Bongardo}, {Aller},
  {Aller}, {De Zotti}, {L{\"a}hteenmaki}, {Longo}, {Milano}, {Tagliaferri},
  {Ter{\"a}sranta}, {Tornikoski}, \& {Urpo}}]{cia04}
{Ciaramella}, A., {Bongardo}, C., {Aller}, H.~D., {et~al.} 2004, \aap, 419, 485

\bibitem[{{Edelson} \& {Krolik}(1988)}]{ede88}
{Edelson}, R.~A. \& {Krolik}, J.~H. 1988, \apj, 333, 646

\bibitem[{{Gu} {et~al.}(2006){Gu}, {Lee}, {Pak}, {Yim}, \& {Fletcher}}]{gu06}
{Gu}, M., {Lee}, C.-U., {Pak}, S., {Yim}, H.~S., \& {Fletcher}, A.~B. 2006,
  \aap, 450, 39

\bibitem[{{Hagen-Thorn} {et~al.}(2004){Hagen-Thorn}, {Larionov}, {Larionova},
  {Kudryavtseva}, {Tikhonov}, {Hagen-Thorn}, {Arkharov}, {di Paola}, \&
  {D'Alessio}}]{hag04}
{Hagen-Thorn}, V.~A., {Larionov}, V.~M., {Larionova}, E.~G., {et~al.} 2004,
  Astronomy Letters, 30, 209

\bibitem[{{Hufnagel} \& {Bregman}(1992)}]{huf92}
{Hufnagel}, B.~R. \& {Bregman}, J.~N. 1992, \apj, 386, 473

\bibitem[{{Korolkov} \& {Parijskij}(1979)}]{kor79}
{Korolkov}, D.~V. \& {Parijskij}, Y.~N. 1979, \skytel, 57, 324

\bibitem[{{Kovalev} {et~al.}(1999){Kovalev}, {Nizhelsky}, {Kovalev}, {Berlin},
  {Zhekanis}, {Mingaliev}, \& {Bogdantsov}}]{kov99}
{Kovalev}, Y.~Y., {Nizhelsky}, N.~A., {Kovalev}, Y.~A., {et~al.} 1999, \aaps,
  139, 545

\bibitem[{{Marscher} {et~al.}(2008){Marscher}, {Jorstad}, {D'Arcangelo},
  {Smith}, {Williams}, {Larionov}, {Oh}, {Olmstead}, {Aller}, {Aller},
  {McHardy}, {L{\"a}hteenm{\"a}ki}, {Tornikoski}, {Valtaoja}, {Hagen-Thorn},
  {Kopatskaya}, {Gear}, {Tosti}, {Kurtanidze}, {Nikolashvili}, {Sigua},
  {Miller}, \& {Ryle}}]{mar08}
{Marscher}, A.~P., {Jorstad}, S.~G., {D'Arcangelo}, F.~D., {et~al.} 2008, \nat,
  452, 966

\bibitem[{{Miller} \& {Hawley}(1977)}]{mil77}
{Miller}, J.~S. \& {Hawley}, S.~A. 1977, \apjl, 212, L47

\bibitem[{{Ostorero} {et~al.}(2004){Ostorero}, {Villata}, \& {Raiteri}}]{ost04}
{Ostorero}, L., {Villata}, M., \& {Raiteri}, C.~M. 2004, \aap, 419, 913

\bibitem[{{Papadakis} {et~al.}(2007){Papadakis}, {Villata}, \&
  {Raiteri}}]{pap07}
{Papadakis}, I.~E., {Villata}, M., \& {Raiteri}, C.~M. 2007, \aap, 470, 857

\bibitem[{{Peterson} {et~al.}(1998){Peterson}, {Wanders}, {Horne}, {Collier},
  {Alexander}, {Kaspi}, \& {Maoz}}]{pet98}
{Peterson}, B.~M., {Wanders}, I., {Horne}, K., {et~al.} 1998, \pasp, 110, 660

\bibitem[{{Raiteri} {et~al.}(2003){Raiteri}, {Villata}, {Tosti}, {Nesci},
  {Massaro}, {Aller}, {Aller}, {Ter{\"a}sranta}, {Kurtanidze}, {Nikolashvili},
  {Ibrahimov}, {Papadakis}, {Krichbaum}, {Kraus}, {Witzel}, {Ungerechts},
  {Lisenfeld}, {Bach}, {Cim{\`o}}, {Ciprini}, {Fuhrmann}, {Kimeridze},
  {Lanteri}, {Maesano}, {Montagni}, {Nucciarelli}, \& {Ostorero}}]{rai03}
{Raiteri}, C.~M., {Villata}, M., {Tosti}, G., {et~al.} 2003, \aap, 402, 151

\bibitem[{{Ravasio} {et~al.}(2002){Ravasio}, {Tagliaferri}, {Ghisellini},
  {Giommi}, {Nesci}, {Massaro}, {Chiappetti}, {Celotti}, {Costamante},
  {Maraschi}, {Tavecchio}, {Tosti}, {Treves}, {Wolter}, {Balonek}, {Carini},
  {Kato}, {Kurtanidze}, {Montagni}, {Nikolashvili}, {Noble}, {Nucciarelli},
  {Raiteri}, {Sclavi}, {Uemura}, \& {Villata}}]{rav02}
{Ravasio}, M., {Tagliaferri}, G., {Ghisellini}, G., {et~al.} 2002, \aap, 383,
  763

\bibitem[{{Ter{\"a}sranta} {et~al.}(2005){Ter{\"a}sranta}, {Wiren}, {Koivisto},
  {Saarinen}, \& {Hovatta}}]{ter05}
{Ter{\"a}sranta}, H., {Wiren}, S., {Koivisto}, P., {Saarinen}, V., \&
  {Hovatta}, T. 2005, \aap, 440, 409

\bibitem[{{Villata} \& {Raiteri}(1999)}]{vil99}
{Villata}, M. \& {Raiteri}, C.~M. 1999, \aap, 347, 30

\bibitem[{{Villata} {et~al.}(2004{\natexlab{a}}){Villata}, {Raiteri}, {Aller},
  {Aller}, {Ter{\"a}sranta}, {Koivula}, {Wiren}, {Kurtanidze}, {Nikolashvili},
  {Ibrahimov}, {Papadakis}, {Tosti}, {Hroch}, {Takalo}, {Sillanp{\"a}{\"a}},
  {Hagen-Thorn}, {Larionov}, {Schwartz}, {Basler}, {Brown}, \&
  {Balonek}}]{vil04b}
{Villata}, M., {Raiteri}, C.~M., {Aller}, H.~D., {et~al.} 2004{\natexlab{a}},
  \aap, 424, 497

\bibitem[{{Villata} {et~al.}(2004{\natexlab{b}}){Villata}, {Raiteri},
  {Kurtanidze}, {Nikolashvili}, {Ibrahimov}, {Papadakis}, {Tosti}, {Hroch},
  {Takalo}, {Sillanp{\"a}{\"a}}, {Hagen-Thorn}, {Larionov}, {Schwartz},
  {Basler}, {Brown}, {Balonek}, {Ben{\'{\i}}tez}, {Ram{\'{\i}}rez}, {Sadun},
  {Boltwood}, {Carini}, {Barnaby}, {Coloma}, {Ros}, {Dai}, {Xie}, {Mattox},
  {Rodriguez}, {Asfandiyarov}, {Atkerson}, {Beem}, {Bloom}, {Chanturiya},
  {Ciprini}, {Crapanzano}, {de Diego}, {Efimova}, {Gardiol}, {Guerra},
  {Kahharov}, {Kapanadze}, {Karttunen}, {Kato}, {Kimeridze}, {Kudryavtseva},
  {Lainela}, {Lanteri}, {Larionova}, {Maesano}, {Marchili}, {Massone},
  {Monroe}, {Montagni}, {Nesci}, {Nilsson}, {Noble}, {Nucciarelli}, {Ostorero},
  {Papamastorakis}, {Pasanen}, {Peters}, {Pursimo}, {Reig}, {Ryle}, {Sclavi},
  {Sigua}, {Uemura}, \& {Wills}}]{vil04a}
{Villata}, M., {Raiteri}, C.~M., {Kurtanidze}, O.~M., {et~al.}
  2004{\natexlab{b}}, \aap, 421, 103

\bibitem[{{Villata} {et~al.}(2002){Villata}, {Raiteri}, {Kurtanidze},
  {Nikolashvili}, {Ibrahimov}, {Papadakis}, {Tsinganos}, {Sadakane}, {Okada},
  {Takalo}, {Sillanp{\"a}{\"a}}, {Tosti}, {Ciprini}, {Frasca}, {Marilli},
  {Robb}, {Noble}, {Jorstad}, {Hagen-Thorn}, {Larionov}, {Nesci}, {Maesano},
  {Schwartz}, {Basler}, {Gorham}, {Iwamatsu}, {Kato}, {Pullen},
  {Ben{\'{\i}}tez}, {de Diego}, {Moilanen}, {Oksanen}, {Rodriguez}, {Sadun},
  {Kelly}, {Carini}, {Miller}, {Catalano}, {Dultzin-Hacyan}, {Fan}, {Ishioka},
  {Karttunen}, {Kein{\"a}nen}, {Kudryavtseva}, {Lainela}, {Lanteri},
  {Larionova}, {Matsumoto}, {Mattox}, {Montagni}, {Nucciarelli}, {Ostorero},
  {Papamastorakis}, {Pasanen}, {Sobrito}, \& {Uemura}}]{vil02}
{Villata}, M., {Raiteri}, C.~M., {Kurtanidze}, O.~M., {et~al.} 2002, \aap, 390,
  407

\bibitem[{{Zhang} {et~al.}(2004){Zhang}, {Zhang}, {Zhao}, {Xie}, {Wu}, \&
  {Zheng}}]{zha04}
{Zhang}, X., {Zhang}, L., {Zhao}, G., {et~al.} 2004, \aj, 128, 1929

\end{thebibliography}
\end{document}